\documentclass{aastex} 

\slugcomment{Submitted to ApJ} 
 
\shorttitle{Raman Scattered He~II in NGC~6790} 
\shortauthors{Kang et al.} 
 
\begin{document} 
 
\title{Raman Scattered He~II~$\lambda$~6545
in the Young and Compact Planetary Nebula NGC~6790} 
 
\author{Eun-Ha Kang$^1$, Byeong-Cheol Lee$^2$ \& Hee-Won Lee$^1$}
\affil{$^1$ Department of Astronomy and Space Science,
Astrophysical Research Center for the Structure and Evolution 
of the Cosmos, Sejong University, Seoul, 143-747, Korea \\
$^2$ Department of Astronomy and Atmospheric Sciences,
Kyungpook National University}
\email{hwlee@sejong.ac.kr}

\begin{abstract} 
We present the high resolution spectra of the young and compact
planetary nebula NGC~6790 obtained with the 
echelle spectrograph at Bohyunsan Optical Astronomy Observatory
and report the discovery of Raman scattered He~II~$\lambda$~6545 
in this object.  This line feature is formed in a thick neutral
region surrounding the hot central star, where He II$\lambda$ 1025 line
photons are scattered inelastically by hydrogen atoms.
A Monte Carlo technique is adopted to compute the line profiles with a 
simple geometric model, in which the neutral region is
in the form of a cylindrical shell that is expanding from the central star.
From our line profile analysis, 
the expansion velocity of the H~I region lies in the range 
$v_{exp}=15-19{\rm\ km\ s^{-1}}$. Less stringent 
constraints are put on the H~I column density $N_{HI}$ and covering 
factor $C$, where the total flux of 
Raman He~II$\lambda$6545 is consistent with their product $CN_{HI}\sim 0.5\times 10^{20}{\rm\ cm^{-2}}$. 
The Monte Carlo profiles from stationary emission models exhibit 
deficit in the wing parts. A much better fit is obtained when the He~II 
emission region is assumed to take the form of a ring that slowly rotates 
with a rotation speed $\sim 18{\rm\ km\ s^{-1}}$.  Brief discussions 
are presented regarding the mass loss processes and future observations.

\end{abstract} 
 
\section{Introduction} 

Mass loss is an important process that mainly occurs in the late stage 
of stellar evolution.  A star with a mass less than $8{\rm\ M_\odot}$
loses a significant amount of mass in the giant stage
before becoming a planetary nebula with a hot white dwarf at its
center. Considering the Chandrasekhar limit
of $1.4 {\rm\ M_\odot}$, the mass loss process in the giant stage 
with enriched heavy elements should be important in the chemical 
evolution of the interstellar medium.
In this regard, with a recent history of mass loss, young planetary
nebulae are interesting objects to study the mass loss process.

It is expected that around a young planetary nebula there may be
a significant amount of neutral material that was lost in the previous 
stage of stellar evolution. In this case, the neutral region is exposed
to the strong UV emission line source in the vicinity
of the hot central star of the planetary nebula. Therefore, important
information related with the mass loss process can be gathered 
from investigations of the scattering processes of the UV radiation
originating from the center region.

Taylor, Gussie \& Pottasch (1990) performed H~I 21 cm radio observations for
a number of compact planetary nebulae (see also Altschuler et al. 1986,
Gussie \& Taylor 1995, Schneider et al. 1987).
Their target selection was made on the basis of high radio brightness
temperature, which is indicative of the nebular compactness.
They searched an absorption trough that may be formed at the radial 
velocity of a compact planetary nebula when the neutral region blocks
the background H~I radio emission from our Galaxy. A number of compact
young planetary nebulae including IC~5117 and NGC~6790 have been detected.
Adopting an excitation temperature $T_{HI}=100{\rm\ K}$, the typical 
H~I column density was determined to be of order 
$N_{HI}\sim 10^{20}{\rm\ cm^{-2}}$ in these objects.

Astrophysical Raman spectroscopy involving atomic hydrogen was initiated 
by Schmid (1989), 
who identified the mysterious broad emission bands occurring at
6825 \AA\ and 7088 \AA\ in many symbiotic stars
(see also Nussbaumer, Schmid \& Vogel 1989).  He proposed that 
a hydrogen atom in the ground state is excited with the absorption 
of an incident far UV O~VI$\lambda$ 1032 photon and de-excites into
the $2s$ level with the re-emission of an optical photon at 6825 \AA.
An analogous process for far UV O~VI$\lambda$1038 yields optical photons
at 7088 \AA. The large line width and prominent linear polarization exhibited
by these scattered features strongly support his proposal (e.g. Harries
\& Howarth 1996).
Observations made simultaneous in the UV and optical regions also
confirm the Raman scattering nature (Espey et al. 1995).

In the spectrum of the symbiotic star RR~Telescopii, Van Groningen (1993) 
discovered Raman scattered He~II features that are formed blueward of hydrogen 
Balmer emission lines. He~II emission lines arising from transitions 
between $n=2k$ and $n=2$ levels have wavelengths that are slightly shorter 
than hydrogen Lyman lines owing to the fact that He~II ions are single 
electron atoms with a slightly larger two body reduced mass. 
The proximity to resonance is responsible for
a large scattering cross section requiring the existence of a neutral
region with $N_{HI}\sim 10^{20}{\rm\ cm^{-2}}$ around a He~II emission 
source. Raman scattered He~II features are also reported in other symbiotic
stars including He 2-106, HM~Sagittae and V1016~Cygni 
(Lee, Kang \& Byun 2001, Jung \& Lee 2004b, Birriel 2004).

Raman scattering of He~II by atomic hydrogen 
also operates in young planetary nebulae. 
The first discovery was reported by P\'equignot et al. (1997) 
in their spectroscopic analysis of the young planetary nebula NGC~7027.
Subsequently, Groves et al. (2002) found the same Raman scattered He~II
features in the planetary nebula NGC 6302. Recently, Lee et al. (2006) 
reported that the compact planetary nebula IC 5117 also exhibits 
Raman scattered He~II features blueward of H$\alpha$ and H$\beta$.
In these objects, it appears that the central star is surrounded by
a neutral region with a significant covering factor. In particular,
Lee et al. (2006) discussed in detail the atomic physics
of He~II recombination and Raman scattering processes. 

We present our high resolution spectra of the young and compact planetary
nebula NGC~6790 and report our finding of the Raman scattered 
He~II$\lambda$6545 feature in this object.  
Using the H$\alpha$ image, Tylenda et al. (2003) measured the angular size 
of NGC~6790 to be $4''\times 3''$. This size estimate of NGC~6790 is 
consistent with the HST image shown by Kwok, Su \& Sahai (2003), who also 
identified two inner shells of similar orientations
in NGC~6790. The distance to NGC~6790 is poorly known. 
Gathier et al. (1986) proposed that NGC~6790
is further than $\sim 0.8 {\rm\ kpc}$ based on their kinematic considerations.
%Sabbadin (1984) proposed a distance to NGC~6790 to be 1.8 kpc. 
Adopting a statistical method Zhang (1995) suggested a distance of 5.7 kpc 
to NGC~6790. In their high resolution spectroscopy of NGC~6790,
Aller, Hyung \& Feibelman (1996) proposed a core mass of 
$0.6{\rm\ M_\odot}$ and an age of 6000 yr with the note that these 
values are dependent on the uncertain distance to NGC~6790.

We perform Monte Carlo radiative transfer simulations in order to obtain 
the geometric and kinematic information of the neutral region.
In section 2, we describe our observation and line fitting analyses
and the following section presents our results of 
the Monte Carlo radiative transfer. In the final section, we discuss
briefly our observation and mass loss processes of NGC~6790.

\section{Observation and Analysis}

\subsection{Observation and Data}

We observed the young planetary nebula NGC~6790 on the night of 2008
May 31 using the 1.8 m telescope at Bohyunsan Optical Astronomy 
Observatory (BOAO). The spectrograph that we used is the BOES (BOAO 
Echelle Spectrograph), which is a bench-mounted echelle 
system fed by optical fibers with various diameters. We used the 300 micron 
fiber, which yields the spectral resolution $\sim 30,000$ with the field of 
view of $3''$. The spectral coverage
ranges 3600 \AA\ through 10,500 \AA. We obtained two spectra with
exposure times of 600 s and 7200 s, respectively. A Th-Ar lamp was used for 
wavelength calibraions.
For more detailed information on BOES, one is referred to Kim et al. (2007).   
Standard procedures  using the IRAF packages were followed
to reduce the spectra. 

\begin{figure}
\centering
\plotone{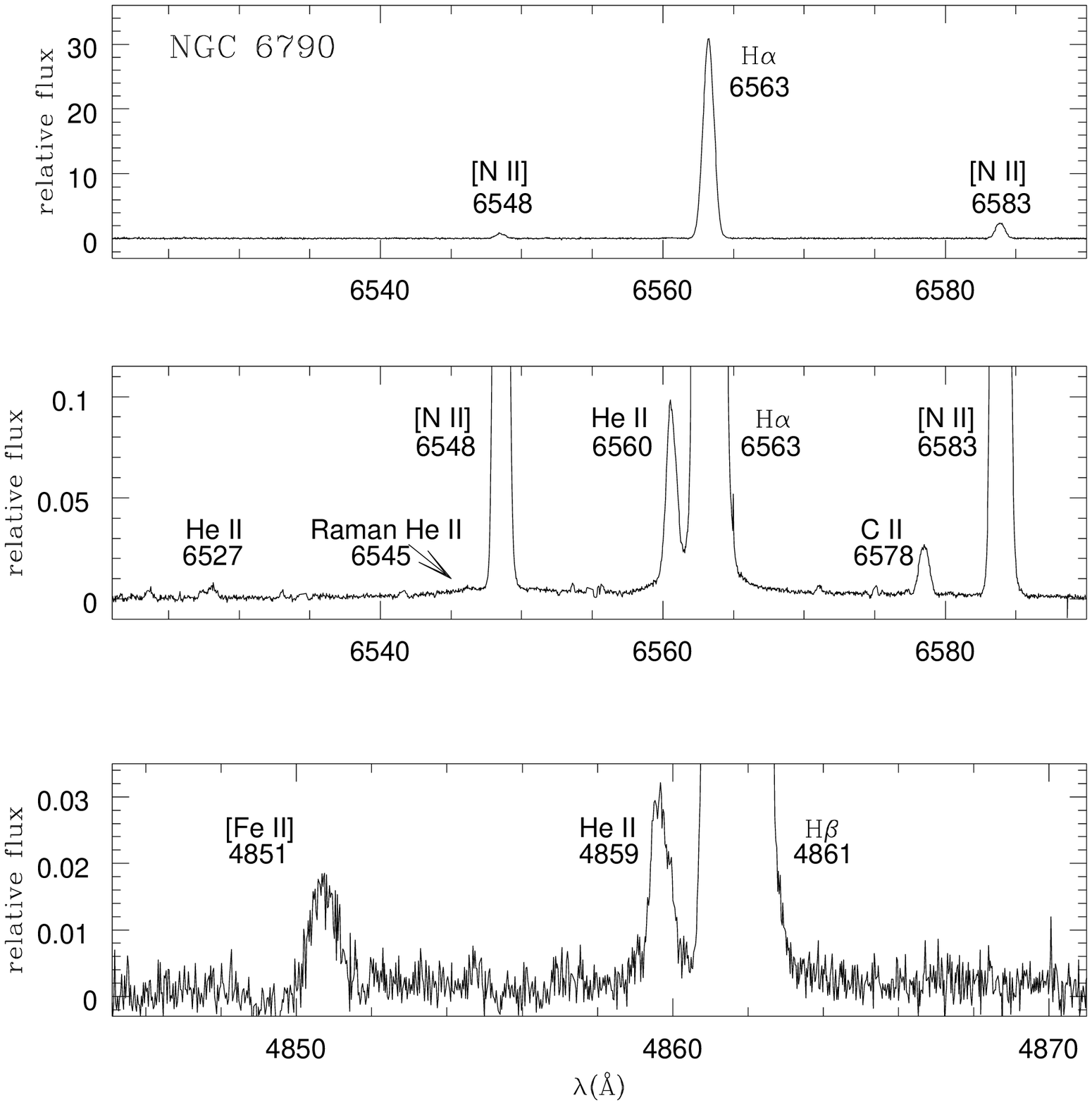}
\caption{High resolution spectra of NGC~6790 obtained with BOES.
The top panel is a short exposure spectrum with exposure time of 600 s,
and the exposure time for the middle and bottom panels is 7200 s. 
The relative flux density is normalized such that [N~II]~$\lambda$6548
has the flux density peak of unity.
In the top panel,
the strongest emission line H$\alpha$ is unsaturated. We also clearly see
[N~II] lines at 6548 \AA\ and 6583 \AA.
In the middle panel, we note that around [N~II]~$\lambda$6548 there exists
a broad wing feature. No similar feature is present around 3 times stronger
[N~II]~$\lambda$6583, which means that the broad wing feature around
[N~II]~$\lambda$6548 is not associated with [N~II] nor is an instrumental
artifacts. The bottom panel shows the H$\beta$ part of the BOES spectrum.
The insufficient quality of the data hinders the clear detection of the Raman
scattered He~II$\lambda$4850.
}
\label{boes1}
\end{figure}

In Fig.~\ref{boes1}, we show parts of our spectra 
around H$\alpha$ and H$\beta$.
The vertical axis represents the relative flux density.
We normalize the flux density using [N~II]$\lambda$6548, which is set 
to have a flux density peak of unity.
The top panel of Fig.~\ref{boes1} is the spectrum around H$\alpha$ 
with an exposure time of 600 s. We note strong forbidden emission lines of
N II at 6548 \AA\  and 6583 \AA. 
In this short exposure spectral image, the strongest H$\alpha$ is
unsaturated, allowing us to fit the H$\alpha$ profile.  
The middle panel of Fig.~\ref{boes1} shows the H$\alpha$ part of the spectrum 
with an exposure time of 7200 s. The strong H$\alpha$ is saturated and we can 
discern very faint emission lines including He~II$\lambda$6527.
He~II$\lambda$6527 arises from transitions between $n=14$ and $n=5$
levels. We also clearly notice that 
around [N II]$\lambda$6548 there exists a broad bump-like feature. 
This feature is not an instrumental artifact because no such feature 
is present near [N II]$\lambda$ 6583, which is supposed to be 3 times 
stronger than [N~II]$\lambda$6548 (e.g. Osterbrock 1987). We propose 
that this broad feature is Raman scattered He~II$\lambda$6545.

The bottom panel of Fig.~\ref{boes1}, we show our spectrum of NGC~6790 
around H$\beta$ with the exposure time of 7200 s. 
If Raman scattered He~II exists blueward of H$\alpha$, we may expect a
similar feature blueward of H$\beta$. Indeed, when P\'equignot et
al. (1997) reported the operation of He II Raman scattering in
NGC~7027, they detected Raman scattered He~II$\lambda$4850. 
In this object, Raman scattered He~II$\lambda$4850 is not blended with other
strong emission lines, which is in contrast with Raman He~II$\lambda$6545 
that is severely blended with [N~II]$\lambda$6548.  In the bottom panel of 
Fig~\ref{boes1}, no broad feature around 4850 \AA\ is detected with a level 
of any significance.  
The quite strong and sharp emission feature at 4851 \AA\ is
an emission line totally irrelevant with Raman scattering. Aller et
al. (1996) identified this emission line as a forbidden line from Fe II.  
Our spectrum is of
insufficient quality to confirm the existence of Raman scattered
He~II$\lambda$ 4850. However, this does
not cast serious doubts of the Raman scattering nature of the 6545
feature, because Raman He~II$\lambda$4850 is always weaker than Raman
He~II$\lambda$6545. More discussion on this point is presented in section 3.2.

\subsection{Line Fitting Analysis}

Single Gaussian functions in the form $f(\lambda)=f_0
\exp[-(\lambda-\lambda_c)^2/\Delta\lambda^2]$ are used to fit the permitted 
emission lines of H$\alpha$, He~II$\lambda$6560, 
He~II$\lambda$6527 and the two N II forbidden lines. The least chi square 
method is adopted to obtain the best fitting Gaussian functions. 
We use the atomic spectral data from the website of the National Institute of 
Standard and Technology(NIST), from which we note that each emission line 
in our spectra of NGC~6790 appears systematically redward of atomic line center
by an amount of 20.7${\rm\ km\ s^{-1}}$.
In Table 1, we summarize the result of our profile analysis.
The fitting parameters are quite similar to those found for IC~5117
by Lee et al. (2006).  

\begin{figure}
\centering
\plotone{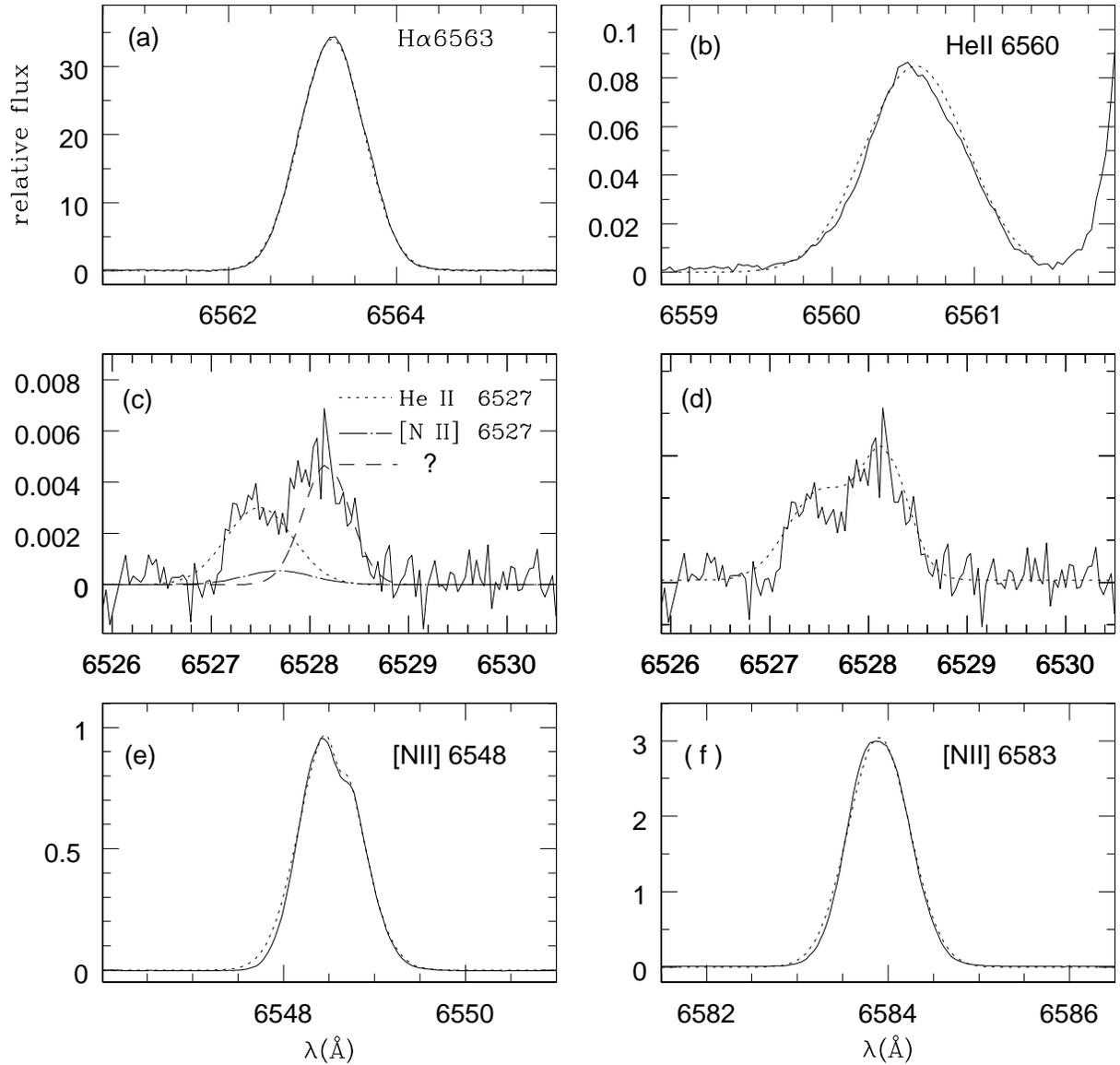}
\caption{Gaussian line fitting analysis. The solid lines show the 
observational data and the dotted
lines show our Gaussian fits. 
The same flux normalization as in Fig.1 is used. 
The top panels show the results for
H$\alpha$ and He~II$\lambda$6560. The middle panels
show the line fitting result for He~II$\lambda$6527 and a nearby
unidentified emission line.  He~II$\lambda$6527 
is fitted by a single Gaussian function, which is shown by the
dotted line in panel (c). The long dashed line in panel (c) shows
the unidentified emission line. Using the atomic data provided by NIST,
we show the line contribution from [N~II]$\lambda$6527 by a dot-dashed line.  
In panel (d) we show the composite
profile from the three single Gaussians shown in panel (c). 
The bottom panels show the detailed views of [N~II] lines.
There is a sharp absorption feature in [N~II]$\lambda$6548, which
is also well fitted by a single Gaussian with the width $\Delta\lambda
=0.1{\rm\ \AA}$ and the center at $\lambda_c=6548.60{\rm\ \AA}$. 
}
\label{linefit}
\end{figure}

Fig.~\ref{linefit} illustrates our line fitting analysis of the emission 
lines in NGC~6790.  The top panels show the result 
for H$\alpha$ and He~II$\lambda$6560. The short exposure data are used 
for the H$\alpha$ emission line, which is excellently fitted by a single 
Gaussian function with a width $\Delta\lambda = 0.54{\rm\ \AA}$.
He~II$\lambda$6560 is also well fitted by a single Gaussian
function with a considerably smaller width of $\Delta\lambda = 0.48{\rm\ \AA}$ 
than that for H$\alpha$.

The middle panels of Fig.~\ref{linefit} show our result for He~II$\lambda$6527,
which is significantly weak compared with He~II$\lambda$~6560.
He~II$\lambda$6527 is strongly blended with another unidentified emission
line.  Because He~II$\lambda$6560 is well-fitted
by a single Gaussian, He~II$\lambda$6527 should be
also fitted by a single Gaussian function, which is shown by the
dotted line in panel (c). The long dashed line in panel (c) shows
our Gaussian fit to the unidentified emission line. 
Groves et al.(2002) noted the existence of [N II]$\lambda$6527 
redward of He II$\lambda$6527 with the wavelength difference of 0.14 \AA\ 
in their spectrum of NGC 6302.
However, the unidentified emission line in our spectrum of NGC 6790 
can not be  [N II]$\lambda$6527, because it appears redward of 
He~II$\lambda$6527 by 1 \AA.
Furthermore, based on the NIST data, [N II]$\lambda$6527 has 
the Einstein A coefficient $A =  5.45\times 10^{-7} s^{-1}$.
Compared with [N II]$\lambda$6548 having $A = 9.19\times 10^{-4}$, 
[N II]$\lambda$6527 should be weaker than [N II]$\lambda$6548 
by a factor of 1700.
Based on these atomic data, we plotted [N~II]$\lambda$6527 
with a dot-dashed line in Fig.2.
As is shown in the figure, [N II]$\lambda$6527 is significantly weaker 
than He~II$\lambda$6527, and hence can not affect the over all line 
fitting result.
The 6528\AA\ feature is much stronger than [N II]$\lambda$6527 and still 
remains to be identified.
Panel (d) shows the composite profiles
of the two single Gaussian functions in panel (c). From our profile
analysis shown also in Table 1, we conclude that the flux ratio of 
He~II$\lambda$~6527 and He~II$\lambda$~6560 is
\begin{equation}
F_{6527}/F_{6560} = 4.1\times 10^{-2} .
\label{heiiratio}
\end{equation}

The bottom panels of Fig.~\ref{linefit} show the detailed profiles of 
[N~II] lines.  It is interesting to note that
[N~II]$\lambda$6548 exhibits a sharp absorption feature centered at 6548.60
\AA. The line center of [N~II]$\lambda$6548 appears at 6548.51 \AA, and the
sharp absorption feature is excellently fitted by a single Gaussian function
with a width of $\Delta\lambda=0.1{\rm\ \AA}$ and center at $\lambda_0
=6548.60{\rm\ \AA}$.  
We find no such absorption feature in [N~II]$\lambda$6583, which
should exhibit exactly the same profile with 3 times more flux (e.g.
Osterbrock 1987).  To our knowledge, no plausible metal transition is 
responsible for this sharp absorption. We also checked the telluric absorption 
lines without finding any strong candidate. In the spectrum of IC~5117
obtained with the 3.6 m Canada-France-Hawaii Telescope
we find no similar absorption feature, for which [N~II]$\lambda$6548
exhibits exactly the same profile as [N~II]$\lambda$6583.  
We tentatively propose that this is attributed
to H$\alpha$ that is redshifted by an amount of 
$v_{abs}\sim 800{\rm\ km\ s^{-1}}$. However, in this work, we limit our 
attention to the Raman 
scattered He~II$\lambda$6545 with no further discussion of this possibly
interesting feature.

Lee et al.(2001) performed a line profile analysis of Raman scattered 
He~II$\lambda$6545 in a number of symbiotic stars. They subtracted one third
of the flux near [N II]$\lambda$6583 from the flux near [N~II]$\lambda$6548 
to expose a broad Raman scattering line feature successfully.
However, in view of the existence of the unidentified absorption feature 
in [N~II]$\lambda$6548 and more severe blending with [N~II]$\lambda$6548, 
we took another approach, in which the Raman scattered He~II$\lambda$6545 
feature is directly fitted from our Monte Carlo data.

\section{Monte Carlo Radiative Transfer}

\subsection{Monte Carlo Procedure}

\begin{figure}
\centering
%\plotone{cylinder.eps}
\plotone{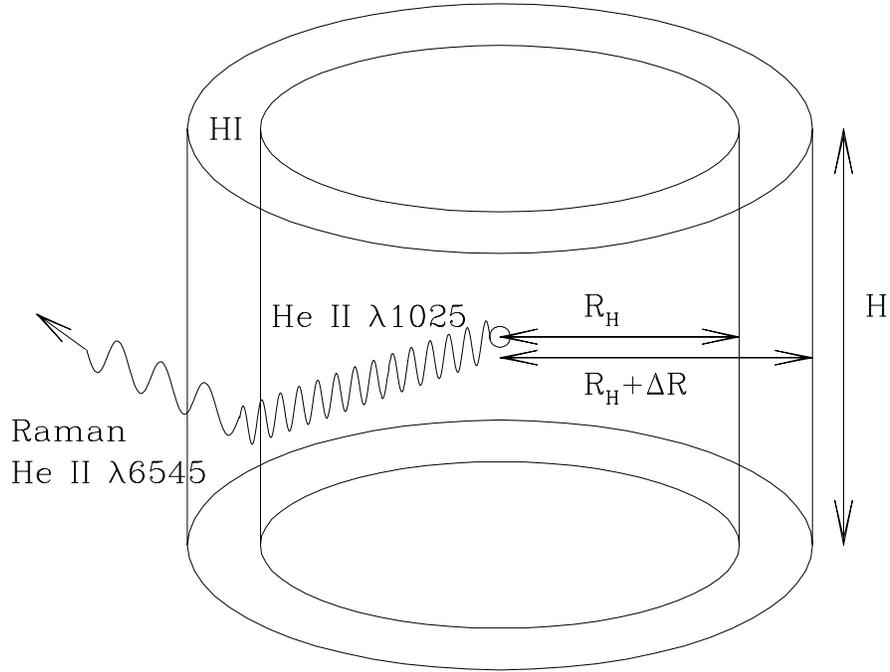}
\caption{
  A schematic diagram of the Raman scattering geometry adopted in this work. 
  The hot star and He~II emission region
  are located at the center.  Surrounding the UV
  emission region, the H~I scattering region takes the form of a 
  cylindrical shell with the inner radius $R_H$, the outer radius 
  $R_H+ \Delta R$ and the  height $H$. In this work, the cylindrical shell 
  is assumed to expand with the speed $v_{exp}$.  Hydrogen atoms are 
  distributed uniformly with a number density $n_H$ inside the cylindrical 
  shell.
}
\label{cylinder}
\end{figure}

In this subsection, we describe the procedure of our Monte Carlo analysis 
of the Raman scattered He~II$\lambda$6545.  
Many planetary nebulae exhibit nonspherical morphology, which may have its
origin in the asymmetric mass loss processes. In the case of NGC~6790,
the HST image obtained by Kwok et al. (2003) shows elongated shells around
the central star. As a first approximation, we adopt a cylindrical shell
model for neutral material, which is schematically illustrated 
in Fig.~\ref{cylinder}.  A similar geometry was considered in the analysis 
of IC~5117 by Lee et al. (2006).

In this cylindrical shell geometry, the hot UV source is located 
at the center and H~I material is uniformly distributed inside the 
cylindrical shell with finite height
and thickness. The same geometry was adopted by Lee et al. (2006).
However, the essential difference is that we now consider the scattering
region is expanding with the constant expansion velocity $v_{exp}$.
The cylindrical region is characterized by a uniform H~I density $n_H$,
the height $H$ and the inner and outer radii $R_H$ and 
$R_H+\Delta R$, respectively. In this case, the H~I column density 
of the cylindrical shell is given by $N_{HI}= n_H \Delta R$. 

Since the shell is of uniform density, instead
of the physical length $l$ we measure
the distance inside the shell in terms of the scattering optical depth $\tau$
defined by
\begin{equation}
\tau = n_H\sigma_{tot} l,
\end{equation}
where $\sigma_{tot}$ is the sum of the cross sections
for Rayleigh and Raman scattering. Since $\sigma_{tot}$ is a sensitive
function of a wavelength of the photon being considered, a given distance
may correspond to different optical depths dependent on the wavelength. 
Therefore, once a photon is generated in the Monte Carlo simulation, we 
assume that
the wavelength does not change as long as it is Rayleigh scattered. 
Considering that the scattering region is neutral, this assumption should
be reasonable.

The basic atomic physics of Raman scattering adopted in our Monte Carlo code 
is explained in detail by Jung \& Lee (2004a). Due to the proximity 
of He II $\lambda$ 1025 to H~I Ly$\beta$ resonance, the scattering cross 
section increases steeply near Ly$\beta$. 
Yoo, Bak \& Lee (2002) showed that the branching ratio $r_b$ 
into Raman scattering increases approximately linearly 
with wavelength, which is given by
\begin{eqnarray}
r_b &=&\sigma_{Ram}/\sigma_{tot} 
\nonumber \\
&=& 0.1342+12.50(\lambda-\lambda_{Ly\beta})/
\lambda_{Ly\beta},
\end{eqnarray}
where $\sigma_{Ram}$ is the cross section for Raman scattering and 
$\lambda_{Ly\beta}$ is the Ly$\beta$ center wavelength. 
Therefore, the Raman conversion into the optical region is quite
sensitive to the incident wavelength, which in turn depends on 
the expansion velocity.

From the energy conservation, a Raman scattered He~II feature is 
characterized by its large width given by 
\begin{equation}
{\Delta\lambda_{Ram}\over\lambda_{Ram}}
=\left({\lambda_{Ram}\over\lambda_i}\right)
{\Delta\lambda_i\over\lambda_i},
\label{linewidth}
\end{equation}
where $\lambda_i$ and $\lambda_{Ram}$ are wavelengths of the incident
and Raman scattered radiation (e.g. Schmid 1989, Nussbaumer et al. 1989).
In the case of Raman He II $\lambda$6545, the profile width becomes
about 6 times broader than He~II$\lambda$1025, which endows a unique 
property that the profile is mainly determined from the relative motion 
between the emitter and the scatterer.

In our Monte Carlo calculation, we also consider the re-entry
of a photon emerging from the inner wall of the cylinder, for which
we assume that this photon travel freely until it hits the inner wall 
on the opposite side. We consider a photon with a unit wavevector 
${\bf\hat k}$ supposed to travel a scattering optical depth $\tau$
from the position ${\bf r}_i=(x_i, y_i, z_i)$. If this photon emerges
from the inner wall of the cylinder, we find the two points of
intersection with the inner wall of the cylinder. This is
accomplished by solving the quadratic for $\tau_p$
\begin{equation}
R_H^2 = |({\bf r}_i+\tau_p{\bf\hat k})\cdot{\hat \rho}|^2,
\end{equation}
for which we denote the two solutions by $\tau_{p1}$ and $\tau_{p2}$ with
$\tau_{p2}>\tau_{p1}$.
Here, $\hat\rho$ is the unit vector pointing radially outward from the
cylinder axis. The difference of the two solutions $\Delta\tau_p$ is
given by  
\begin{eqnarray}
\Delta\tau_p &=& \tau_{p2}-\tau_{p1}
\nonumber \\
&=&{2\sqrt{R_H^2(1-k_z^2)-(k_xy_i-k_yx_i)^2} \over
(1-k_z^2)},
\end{eqnarray}
where $k_x, k_y$ and $k_z$ are the components of ${\bf\hat k}$.
By adding $\Delta\tau_p$ to the original photon path, we find the new 
scattering site in the other side of the shell.

The incident He~II$\lambda$1025 line flux and profile can be inferred
from the case B recombination theory of single electron atoms provided by 
Storey \& Hummer (1995). In Table 2, we show the expected 
He~II$\lambda$1025 line flux relative to He~II$\lambda$6560 and 
He~II$\lambda$6527
for electron number densities $n_e=10^4, 10^6$ and $10^8{\rm\ cm^{-3}}$
and temperatures $T_e=10^4$ and $2\times10^4{\rm\ K}$. 
We note that our observed flux ratio 
of He~II$\lambda$~6527 and He~II$\lambda$~6560 given 
in Eq.~(\ref{heiiratio}) is consistent with the nebular condition 
of $n_e \sim 10^6{\rm\ cm^{-3}}$ 
and $T_e=10^4{\rm\ K}$. However, this choice is not unique and 
the range of He~II$\lambda$1025
is already quite significant with the choice of parameters in Table 2.
With this caveat in mind, we fix the electron number density $n_e=10^6
{\rm\ cm^{-3}}$ and $T_e=10^4{\rm\ K}$.
Adopting these values of $n_e$ and $T_e$, the recombination theory 
by Storey \& Hummer (1995) gives
$F_{1025}=4.2 F_{6560}$, which is used for our Monte Carlo calculations.

The Monte Carlo simulation starts with a generation of He~II$\lambda$1025
line photons having the same line profile with that of observed 
He~II$\lambda$6560, and appropriately scaled using the recombination 
theory.  As He~II$\lambda$6560 is fitted by a single Gaussian
with a width of $\Delta\lambda=0.48{\rm\ \AA}$, we note that the line
profile function $f_{UV}$ for He~II$\lambda$1025 is given by
\begin{equation}
f_{UV}(\lambda)=f_{1025}\exp{-[(\lambda-\lambda_{1025})^2/
\Delta\lambda_{1025}^2}]
\label{he1025}
\end{equation}
with $\Delta\lambda_{1025}=0.48\cdot 1025/6560{\rm\ \AA}=0.075{\rm\ \AA}$.
Here, the peak value $f_{1025}$ is appropriately adjusted to yield
$F_{1025}=4.2 F_{6560}$.

We trace each individual He~II$\lambda$1025 line photon until it escapes
from the H~I region.  From Eq.~(\ref{linewidth}), it is noted that
the profiles of the Raman scattered features are determined from
the relative kinematics between the emission source and the H~I region
and almost independent of the observer's line of sight.
Therefore, in this work, we collect all the photons irrespective
of the final direction. 

\subsection{Simulated Raman Profiles}

\subsubsection{Spherical Emission Region}

In the work of Lee et al. (2006), the analysis of Raman scattered
He~II$\lambda$6545 was purely based on the atomic physics
and focused on the exact location of line center. Their computation
shows that the Raman scattered feature should be centered significantly
blueward of [N~II]$\lambda$6548.  In Fig.~\ref{boes1}, we note that the 
Raman He~II$\lambda$6545 is completely blended with [N~II]$\lambda$6548,
which implies that the neutral scattering region
should be receding from the central UV source. 

\begin{figure}
\centering
\plotone{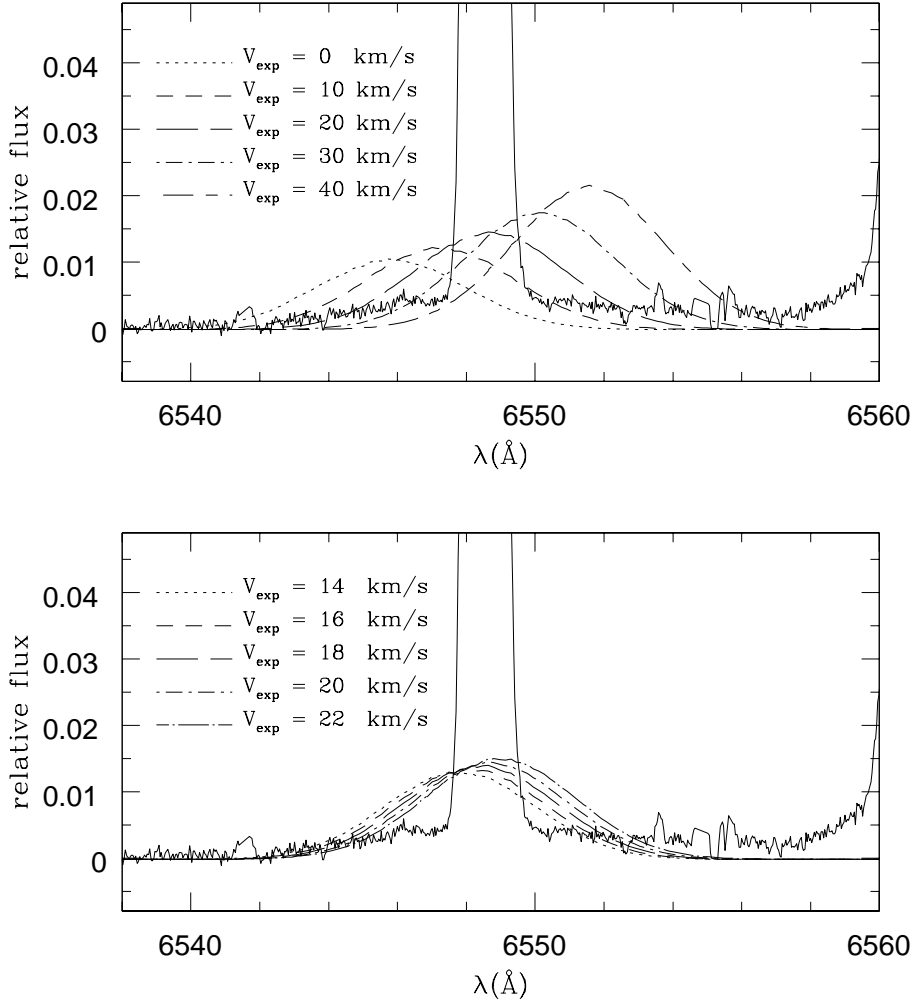}
\caption{
Line profiles of Raman scattered He~II$\lambda$6545 from our Monte Carlo
simulations for various expansion speeds. The covering factor is fixed to
be unity and $N_{HI}=10^{20}{\rm\ cm^{-2}}$. Due to the inelasticity of
Raman scattering or Eq.(\ref{linewidth}), the location of line center
is fairly sensitive to $v_{exp}$. The top panel shows the profiles 
for velocities 
in the range $v_{exp} \le 40  {\rm\ km\ s^{-1}}$ in an interval of $10
{\rm\ km\ s^{-1}}$.  The bottom panel shows the profiles 
for velocities in the range $14{\rm\ km\ s^{-1}} \le v_{exp} \le 
22{\rm\ km\ s^{-1}}$. 
It is notable that the expansion velocity $v_{exp}$ affects both the
location of line center and the total Raman flux. 
}
\label{mc_res1}
\end{figure}

In Fig.~\ref{mc_res1}, we show our Monte Carlo profiles
for various expansion speeds $v_{exp}$ of the neutral scattering
region with respect to the hot central star. 
In this figure, the height of the cylinder is taken to be infinite so that
the covering factor of the scattering region is unity. The column density
is fixed to $N_{HI}=1\times10^{20}{\rm\ cm^{-2}}$. 
The solid line shows our observed data and the other lines show 
our Monte Carlo profiles corresponding to various values of $v_{exp}$.  
We can clearly notice the center shift of the Raman 
He~II$\lambda$6545, which is highly enhanced due to the line broadening given 
in Eq.~(\ref{linewidth}).  The top panel shows the profiles for velocities 
$v_{exp} \le 40  {\rm\ km\ s^{-1}}$.  The bottom panel shows 
the profiles 
for velocities in the smaller range $14{\rm\ km\ s^{-1}} \le v_{exp} \le 
22{\rm\ km\ s^{-1}}$. From the figure, the plausible expansion velocity 
is around $20{\rm\ km\ s^{-1}}$, for which the peak wavelength resides 
inside the [NII]$\lambda$6548 emission line. 

One interesting point to note from Fig.~\ref{mc_res1} is that
the strength of the Raman feature increases sharply as $v_{exp}$ increases
despite the fact that the covering factor and $N_{HI}$ are fixed.
This is explained by the fact that the Raman
scattering cross section sharply increases near H$\alpha$ due to
Ly$\beta$ resonance in the parent wavelength space. Therefore,
a receding H~I region yields more Raman scattered He~II$\lambda$6545
photons than when the same region is stationary.
This complicated dependence of the scattering cross section
on wavelength also results in slightly asymmetric Raman profiles,
which is barely noticeable in Fig.~\ref{mc_res1}.
Therefore, the Raman conversion efficiency may be estimated
accurately only after the kinematics of the scattering region
with respect to the emission source is carefully determined.

\begin{figure}
\centering
\plotone{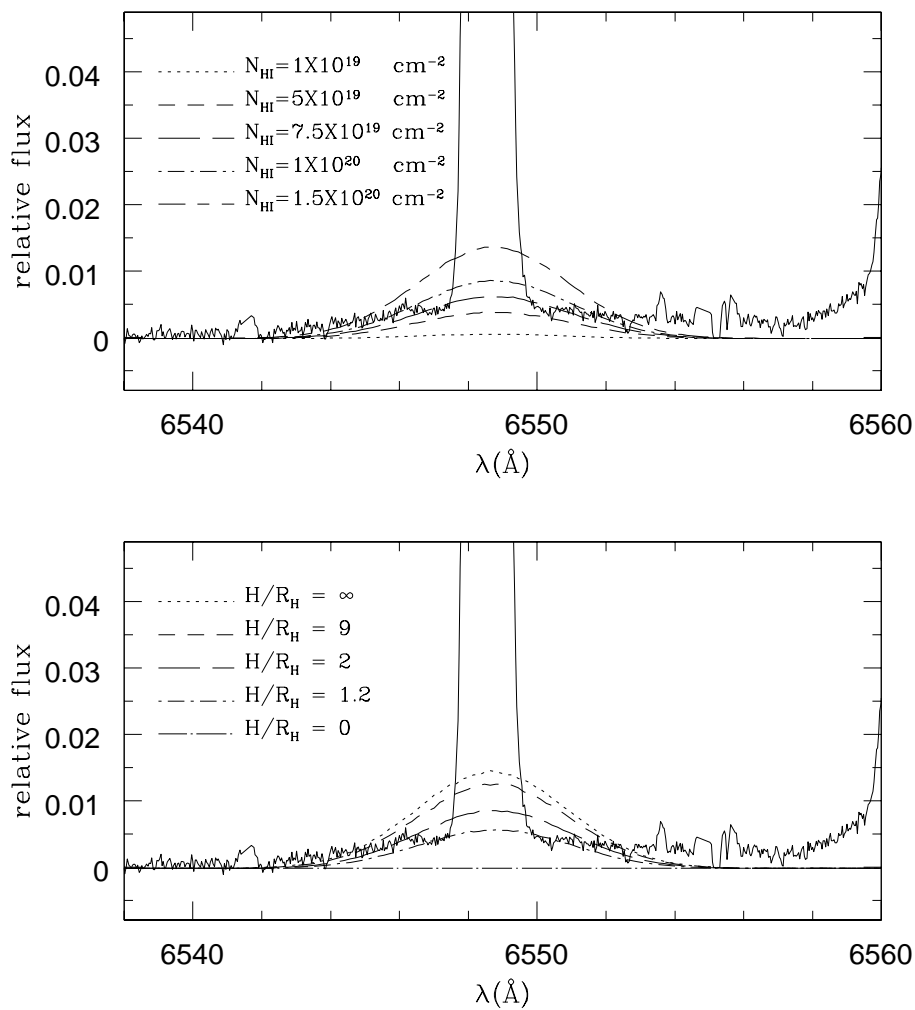}
\caption{
Monte Carlo line profiles of Raman scattered He~II$\lambda$6545 
for various $N_{HI}$ and covering factors. The left panel shows the Monte
Carlo profiles for various $N_{HI}$ with the covering factor fixed to
be $H/R_H=2$. The right panel shows the simulated profiles for
various covering factors with fixed $N_{HI}=10^{20}{\rm\ cm^{-2}}$.
}
\label{mc_res2}
\end{figure}

In the left panel of Fig.~\ref{mc_res2}, we show the Raman 
profiles for various H~I column densities ranging 
$N_{HI}=10^{19}-1.5\times10^{20} {\rm\ cm^{-2}}$ with the fixed 
values of $H/R_H = 2$ and $v_{exp}=20{\rm\ km\ s^{-1}}$. 
Within this range of $N_{HI}$, the overall strength is nearly
proportional to the H~I column density, because the H~I region is
mostly optically thin with respect to Raman scattering of He~II$\lambda$1025.
This expansion speed is very similar to the value of $16{\rm\ km\ s^{-1}}$ 
determined from Doppler shifted Na D absorption lines by 
Dinerstein, Sneden \& Uglum (1995). 

The right panel of Fig.~\ref{mc_res2} shows the Monte Carlo Raman
profiles for various covering factors of the cylindrical shell.
As is expected, the overall strength is also proportional to
the covering factor.  In both the panels of Fig.~\ref{mc_res2}, 
we obtain qualitatively similar profiles.  This implies
that the Raman profile analysis severely suffers from the degeneracy
problem involving the covering factor and H~I column density.  

\begin{figure}
\centering
\plotone{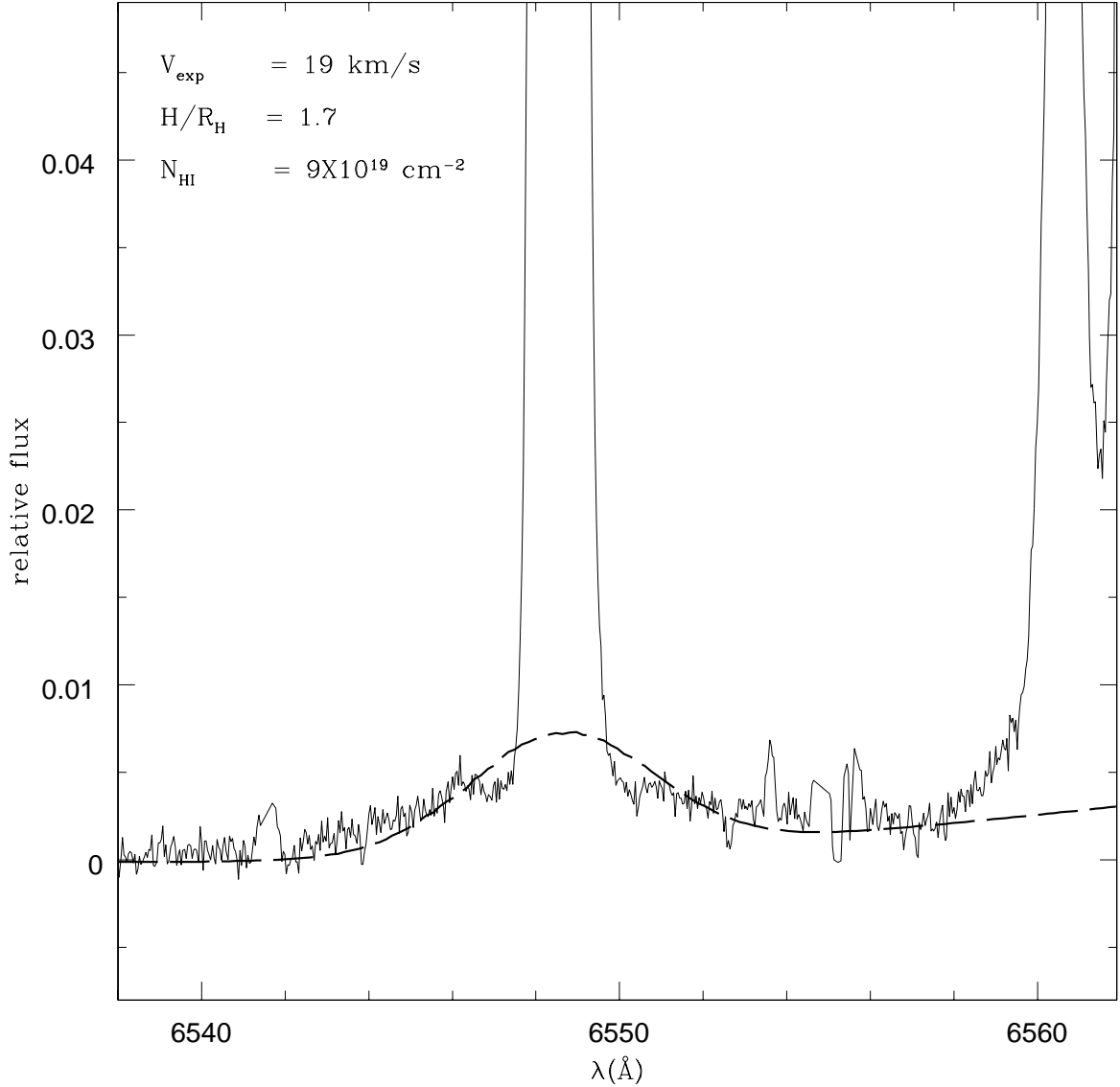}
\caption{
Our best fit Monte Carlo profile of Raman scattered He~II$\lambda$6545 
from a stationary emission region surrounded by a cylindrical shell.
The dotted line is the Monte Carlo line profile and the solid line is the
observed data.  The adopted parameters are $v_{exp}=19{\rm\ km\ s^{-1}}$, 
$H/R_H = 1.7$, $N_{HI}=9\times10^{19}{\rm\ cm^{-2}}$.
}
\label{mc_res3}
\end{figure}

With this caveat in mind related with the degeneracy in $N_{HI}$ and
the covering factor, we show our best fit profile from the Monte Carlo 
calculations in Fig.~\ref{mc_res3}. The model parameters are $v_{exp}=19
{\rm\ km \ s^{-1}}$, $N_{HI}=9\times10^{19}{\rm\ cm^{-2}}$ and 
$H/R_H=1.7$. As in IC~5117,
the H~I region significantly covers the hot central star in NGC~6790.
However, in this figure we
notice that the model profiles exhibit deficit both in the blue wing 
and red wing parts. If this deficit is real, then it implies that
in the direction to the H~I region 
the incident profile is broader than in the observer's line of 
sight. The next subsection discusses this point.

Jung \& Lee (2004b) developed a Monte Carlo code to compute
the line profile of Raman scattered He~II 4850 and analyzed
their spectrum of the symbiotic star V1016~Cyg.
Using the same code, we show in Fig.~7 the Monte Carlo profile
for Raman scattered He~II 4850 by a long dashed line. The same
column density and covering factor as in Fig.~6 were used in this
calculation. In the figure, the solid line shows the BOES data with
the exposure time of 7200 s. Our observational data are barely
consistent with our interpretation of Raman scattering nature.
The poor quality of the current observational data hinders a further
serious quantitative analysis. A more fruitful analysis may
be made only after observational data with a better quality
are secured.

\subsubsection{Ring-like Emission Region}

In this subsection, we perform line profile analyses in the case 
where the emission region
takes the form of a ring that is rotating in the vicinity of the hot central 
star. 
In the previous section, it was assumed that the He~II emission region 
is spherically symmetric and stationary. However, it is highly
probable that the distribution of nebular material significantly deviates 
from spherical symmetry considering the non-spherical shape exhibited by
most planetary nebulae (e.g. Corradi \& Schwarz 1995). 
In this case,
the emission region may plausibly possess an ordered motion component,
which may also be associated with the nonspherical nebular morphology. 
Therefore, we may expect that ionized material
is concentrated on the equatorial plane having some slow rotation
velocity component.

There exists little kinematic information available on the
emission region very near the central star.
No observational data of NGC~6790 are available in the archives of 
HUT and FUSE. 
In consideration of the absence of a unique kinematic model accounting 
for all the observed emission line profiles, we adopt a simple ring-like
emission region, in which we investigate the line of sight effect
on the profiles of the He~II emission and Raman scattered lines.  
Depending on the line of sight of the observer, the rotation velocity 
component is reduced by the factor $\sin i$, where $i$ is the inclination 
angle of the ring.
However, Eq.~(\ref{linewidth}) dictates that the Raman profile is determined
by the velocity component of the emitter with respect to the scatterer
and fairly insensitive to the line of sight.  

This proposition leads to an interesting interpretation of our profile fitting 
of H$\alpha$ and He~II$\lambda$6560 presented in the previous section. 
We may decompose the emission profiles into a bulk component and a random
component. We further assume that the bulk component
represents a slow rotation in the equatorial plane and that the random
component is attributed to a thermal motion and a turbulent motion.
For the sake of simplicity, we assume that He~II$\lambda$6560 and H$\alpha$
are formed in the same ring-like region 
that is in slow rotation in the equatorial plane 
with the speed $v_{bulk}$. 

A He ion being 4 times heavier than a hydrogen 
nucleus, the line width of He~II due to the thermal motion
is half of that for H$\alpha$ if they are formed in the same region. 
However, if the emission region possesses
some turbulent component, then overall random motion component for
hydrogen is broader than that of He~II by a factor less than 2.  
If we denote the electron temperature of NGC~6790 by $T_e=10^4\ T_4{\rm\ K}$, 
then the thermal velocity associated with H$\alpha$ is given by
\begin{equation}
v_{th,H}=\sqrt{k_BT_e\over 2m_p} = 13\ T_4^{1/2}{\rm\ km\ s^{-1}},
\end{equation}
where $m_p$ is the proton mass and $k_B$ is the Boltzmann constant
(e.g. Rybicki \& Lightman 1979).  
Introducing $v_{turb}$ for the turbulent velocity scale, we denote the random
velocity components of H$\alpha$ and He~II$\lambda$6560 
by $v_{ran,H}$ and $v_{ran,He}$, respectively, where
\begin{eqnarray}
v_{ran,H} &=& v_{turb}+v_{th,H} 
\nonumber \\
v_{ran,He} &=& v_{turb}+(v_{th,H}/2).
\end{eqnarray}

Noting that there are three model parameters, namely $i$,
$v_{bulk}$ and $v_{turb}$ for the two line widths, we also encounter a
degeneracy problem. Hoping that future observations may provide independent
constraints on some of these model parameters, we just pick out 
a set of values that yield a reasonable fit to our observed data.  
In the top panels of Fig.~\ref{lineofsight}, we show model line profiles 
for He~II$\lambda$6560 and H$\alpha$ from one such set consisting of 
\begin{eqnarray}
\sin i & =&0.6,\quad v_{bulk} = 18{\rm\ km\ s^{-1}},
\quad v_{turb} = 14{\rm\ km\ s^{-1}}, 
\nonumber \\
v_{th, H} &=& 14{\rm\ km\ s^{-1}},\quad v_{th,He}=0.5 v_{th, H}.
\end{eqnarray}
The thermal velocity $v_{th, H}=14{\rm\ km\ s^{-1}}$ is consistent with
the electron temperature $T_e=10^4{\rm\ K}$, which is similar to that
obtained by Aller et al. (1996) from their photoionization modeling.  
The overall fits to both H$\alpha$ and He~II$\lambda$6560 appear quite 
good. The bulk velocity component is consistent with the size of the 
emission ring region of order $1{\rm\ AU}$ if we interpret the bulk
motion to be Keplerian. However, the bulk motion may not be related
with the Keplerian motion but may be related with the rotation component
of the central star, for which case the physical size of the emission region 
can be at best poorly constrained.

Because the H~I region is also concentrated
on the equatorial plane, the full bulk velocity component 
should be considered without the inclination effect for far 
UV He~II$\lambda$1025 
that is incident on the H~I region.  
In the bottom panel of Fig~\ref{lineofsight}, the dotted line shows 
the He~II$\lambda$6560 profile that would be measured by a hypothetical 
observer 
in the equatorial plane.  It is excellently fitted by a single
Gaussian function with a width $\Delta\lambda = 0.61{\rm\ \AA}$, which is
significantly larger than the observed value of $\Delta\lambda=0.48{\rm\ \AA}$
by a factor of 1.3. Hence, the emission profile for He~$\lambda$1025 incident 
on the neutral region should also be broadened by the same factor.

In Fig.~\ref{final}, we show our Monte Carlo result using the profile
shown in the bottom panel of Fig.~\ref{lineofsight} and appropriately scaled
to He~II$\lambda$1025. The other model
parameters are also adjusted for better fit and they are
$v_{exp}=15{\rm\ km\ s^{-1}}, N_{HI}=9\times10^{19}{\rm\ cm^{-2}}$
and $H/R_H=1.2$. A much better fit is obtained than that considered in
the previous section.  However, it should also be pointed out that 
in constructing the profile in Fig.~\ref{final} 
more model parameters have been used than in the previous section 
and still the degenerate nature of the problem persists.

The expansion velocity of the H~I shell in Fig.~\ref{final} is only 
$v_{exp}=15 {\rm\ km\ s^{-1}}$, which is significantly smaller than 
the value $v_{exp}=19{\rm\ km\ s^{-1}}$ presented in Fig.~6.
This notable discrepancy in expansion velocity is attributed to the
scattering cross section that is sharply peaked around H$\alpha$.
According to Jung \& Lee (2004a), this leads to the center shift of
a Raman scattered He~II feature, which is dependent on the column density.
The result shown in Fig.~\ref{final} implies that
the shape or the width of the incident profile also affects the location
of the line center.  A more quantitative investigation in a significantly 
large parameter space is left to the future work.

\section{Discussion}

H~I Raman spectroscopy provides an accurate determination of the expansion 
velocity of the H~I region, for the measurement of which H~I 21 cm radio 
observation has been the unique tool so far. Our analysis shows that
the expansion velocity lies between $15{\rm\ km\ s^{-1}}$ and 
$19{\rm\ km\ s^{-1}}$, which is consistent with the value of
$16{\rm\ km\ s^{-1}}$ provided by Taylor et al. (1990).
As was pointed out by Lee et al. (2006) the Raman spectroscopy 
allows one to determine the H~I column density 
whereas the excitation temperature should be assumed before
$N_{HI}$ is deduced from H~I 21 cm radio observation. 
According to Taylor et al. (1990), $N_{HI}=2.7\times
10^{20}{\rm\ cm^{-2}}$ assuming the excitation temperature $T_{HI}=
100{\rm\ K}$. Our Raman profile analysis lends support to this excitation
temperature.

Our current data are of insufficient quality 
to lift the degeneracy of the covering factor and H~I column density,
and the overall strength of the Raman feature is determined from
the product of the two quantities. However, our Monte Carlo
calculations show that Raman profiles exhibit redward asymmetry due
to enhanced scattering cross section toward H$\alpha$ resonance.
With better quality spectra that may be available from bigger telescopes,
it is hoped that tighter constraints
are obtained from more refined profile analyses.
If Raman scattered He II 4850 blueward of H$\beta$ can also be used,   
additional constraints can be put to break the degeneracy.

Even though the distance to NGC~6790 is highly uncertain, 
we may assume that the distance is about 1 kpc for simple order of
magnitude calculations.  According to Tylenda et al. (2003),
the angular size of NGC~6790 is $\sim
3''$. This gives a physical size of the H~I region $R\sim 5\times10^{16}
{\rm\ cm}$. If the H~I region is of a thin cylindrical shell with the
height similar to its radius, the total number $N_{tot}$ of 
hydrogen atoms inside the shell is approximately given by
$ N_{tot} = 2\pi R^2 N_{HI} \sim 6\times 10^{54}$.
Here, in our order of magnitude estimate, we ignore the inclination effect, 
which will overestimate the total number of hydrogen atoms 
by the factor $\sin i$. The H~I mass of the neutral region is inferred to be
$M_{HI}\sim 4\times10^{-3}{\rm\ M_\odot}$.

Furthermore, the expansion velocity of $v_{exp}\sim 15{\rm\ km\ s^{-1}}$
and the physical size of $R\sim 5\times 10^{16}{\rm\ cm}$ 
together imply the age of order of a thousand years for NGC~6790.
It should be pointed out that these rough calculations 
are highly dependent on the assumed distance to NGC~6790 and still
the physical size of the H~I region is quite uncertain.

The origin of sharp absorption feature that appeared in [N II]$\lambda$6548 
is quite uncertain. If this absorption feature is attributed to H$\alpha$, 
then it may imply the existence of clumpy components having a small covering 
factor with respect to the [N~II] emission region and
receding with a significant velocity of $\sim 800{\rm\ km\ s^{-1}}$. 
In some planetary nebulae including M2-9 and NGC~6543, it is known that 
fast collimated outflows exist around the central star with a velocity of
order $1000{\rm\ km \ s^{-1}}$ (Balick 1989, Gruendl, Chu \& Guerrero 2004, 
Prinja et al. 2007). Ueta, Fong \& Meixner (2001) presented near IR 
imaging observations of AFGL~618 and 
reported their findings of molecular bullet-like features
moving faster than $200{\rm\ km\ s^{-1}}$.  
However, it still remains a mystery 
whether a clumpy bullet-like object can be ejected with so large 
a velocity from the center region.

It should be pointed out that a ring-like emission
model may not be a unique choice for the observed profiles of 
He~II$\lambda$6560 and H$\alpha$. Many kinematical models involving
jet-like outflows or radial infall and/or outflows may also yield
similarly well-fitting profiles. Therefore without convincing support
from other studies such as imaging observations using interferometry
or hydrodynamical computations, it appears to be too early to conclude
about the kinematics of the He~II emission region.  

A ring-like emission region and H~I region concentrated in the equatorial
region may provide interesting opportunities for spectropolarimetry.
In symbiotic stars, Raman scattered O~VI$\lambda\lambda$6825, 7088
are known to exhibit strong linear polarization (e.g. Harries \&
Howarth 1996, Schmid 1998). The polarization structure may be closely
related with the accretion and mass loss processes
that deviate from spherical symmetry (e.g. Lee \& Park 1999, 
Lee \& Kang 2007, Ikeda et al. 2004). Because Raman scattered features 
consist of purely scattered 
photons, they make ideal targets for linear spectropolarimetry. Future
spectropolarimetric studies may provide more interesting information
regarding the mass loss processes in AGB stars and planetary nebulae.

\acknowledgments 
 
We are grateful to the staffs at the Bohyunsan Optical Astronomy 
Observatory. We also thank an anonymous referee for the constructive
comments, which significantly improved the presentation of our work.
 This research was supported by the Astrophysical Research 
Center for the Structure and Evolution of the Cosmos (ARCSEC'') funded 
by the Korea Science and Engineering Foundation.
 
%\appendix 
% 
%\section{Appendicial material} 
% 

%%% Figures

\begin{figure}
\centering
\plotone{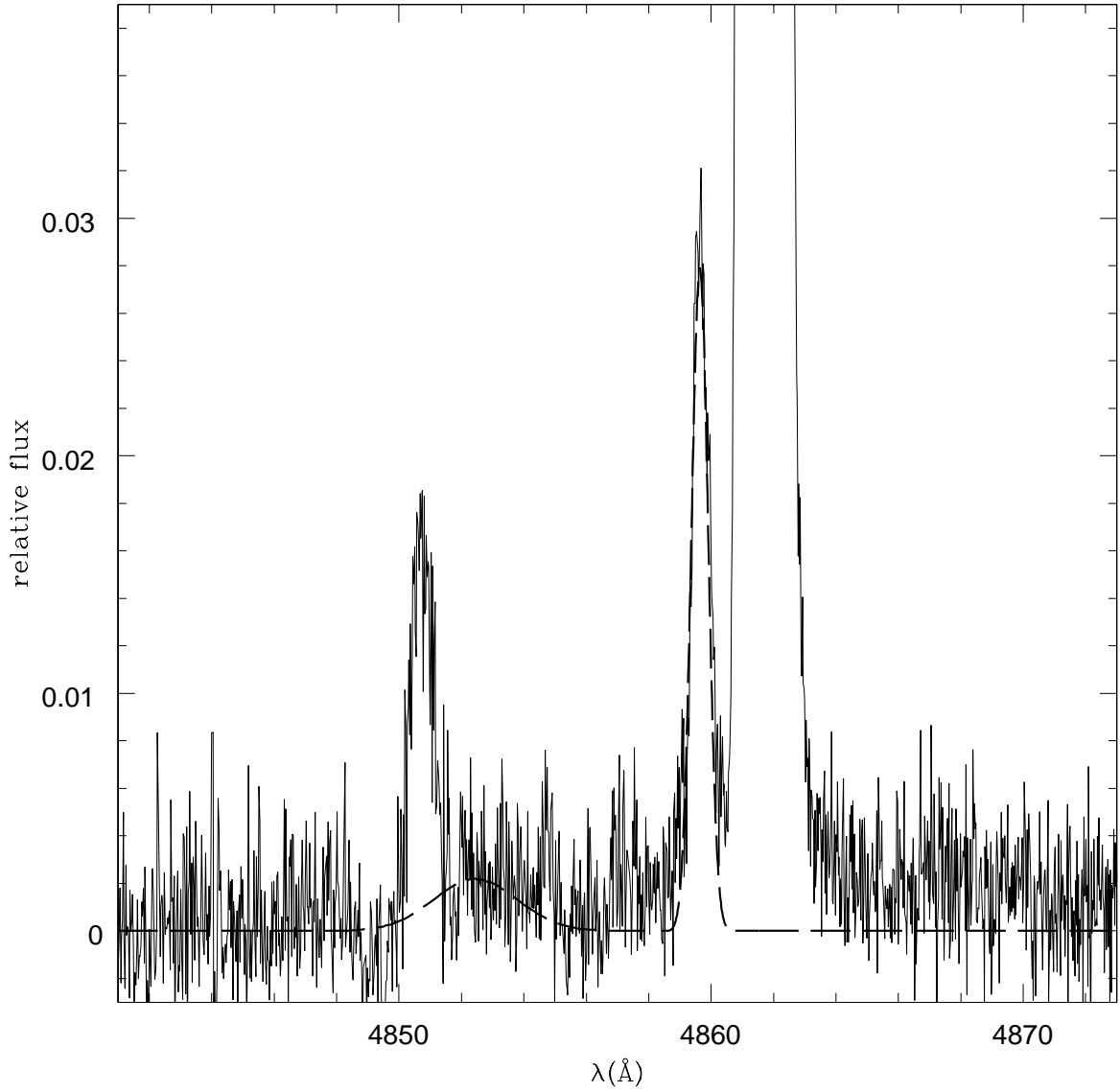}
\caption{
BOES data around H$\beta$(solid line)
and the Monte Carlo profile of Raman scattered He~II$\lambda$4850
(long dashed line). The same column density and covering factor as in Fig.6 
were used in the Monte Carlo calculation. The observational data are barely
consistent with the Monte Carlo result.
}
\label{lineofsight}
\end{figure}

\begin{figure}
\centering
\plotone{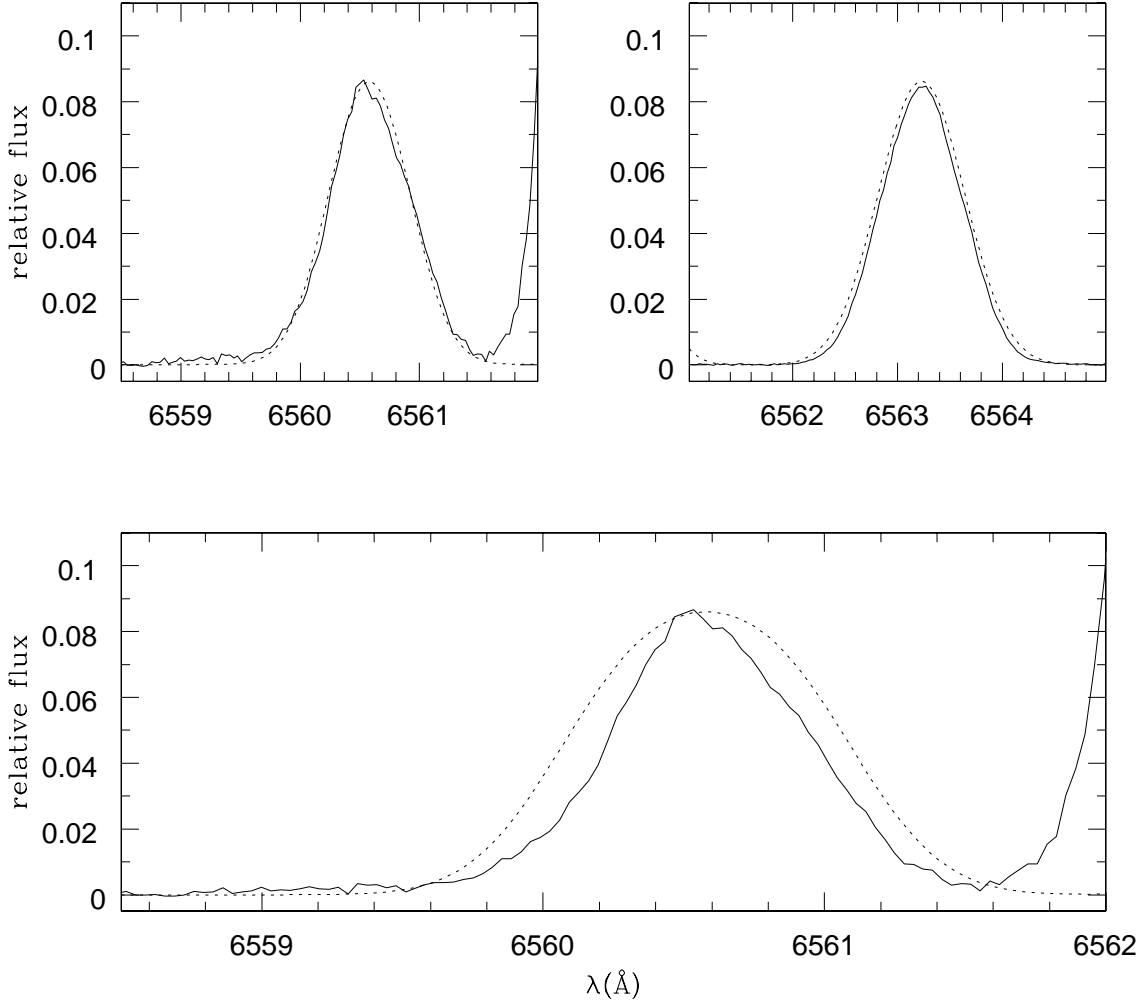}
\caption{
Line profiles of He~II$\lambda$6560 and H$\alpha$
from a ring-like emission region. The axis of the ring makes
an angle $i$ with the line of sight, where we take $\sin i=0.6$ as
an example.  The upper panels show line profiles of He~II$\lambda$6560
and H$\alpha$ viewed from the observer's line of sight. 
The fitting parameters are $v_{bulk}=18{\rm\ km\ s^{-1}}, 
v_{turb}=14{\rm\ km\ s^{-1}}$ and $v_{th,H}=14{\rm\ km\ s^{-1}},
v_{th,He}=0.5v_{th,H}$. See the text of the definitions of these
velocities.  The solid
lines represent the BOES data and the dotted lines are model profiles.
The lower panel shows the observed He~II$\lambda$6560 profile (solid
line) and the model profile that would be observed in the equatorial
direction. The dotted model profile is excellently fitted by a single 
Gaussian with a width $\Delta\lambda=0.61{\rm\ \AA}$.
}
\label{lineofsight}
\end{figure}

\begin{figure}
\centering
\plotone{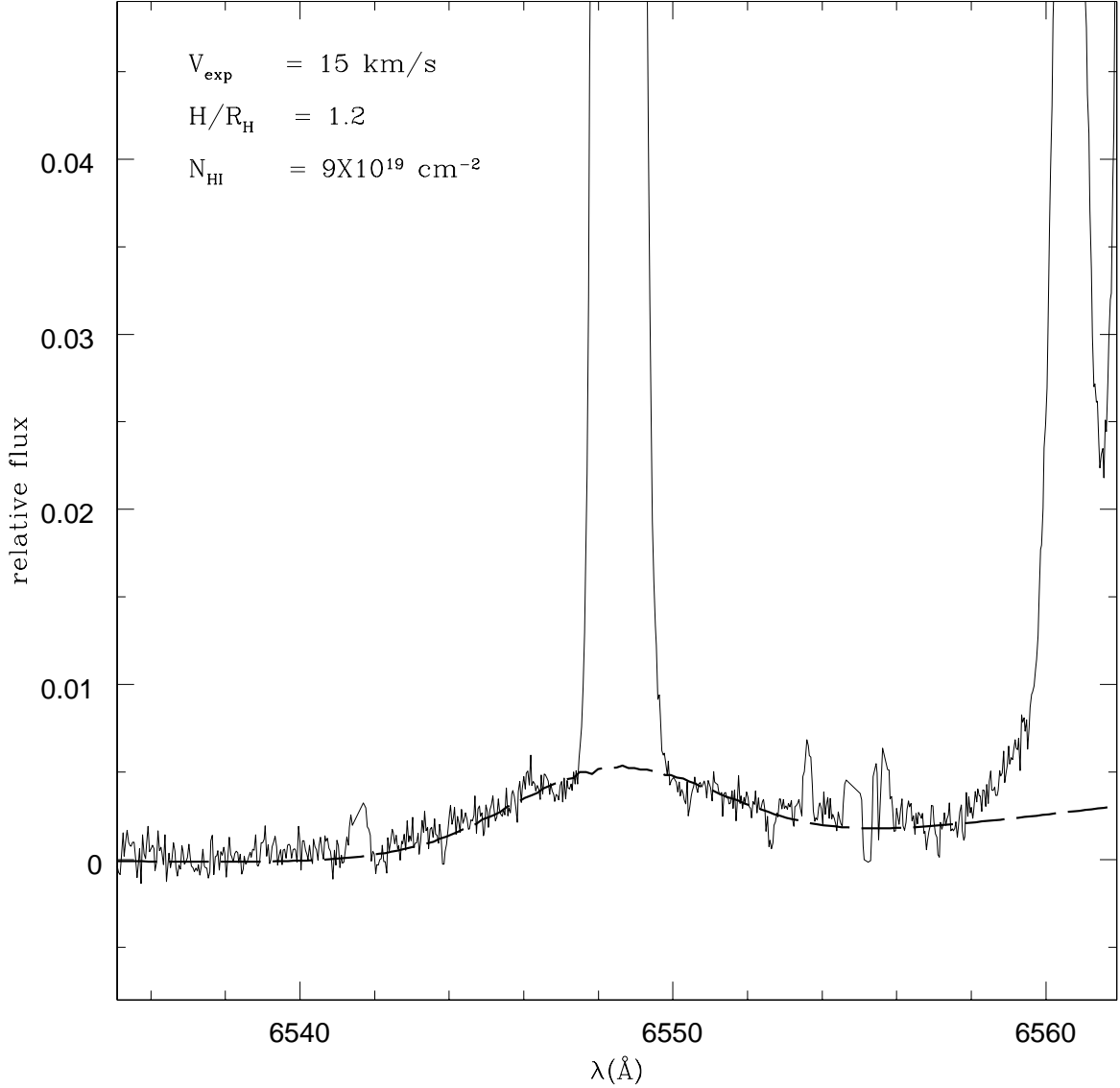}
\caption{
A Monte Carlo best fit profile (dotted line) of Raman scattered 
He~II$\lambda$6545 from a ring-like emission region considered in Fig.~7.
The adopted model parameters are
$\sin i  =0.6$, $v_{bulk} = 18{\rm\ km\ s^{-1}}$,
$v_{turb} = 14{\rm\ km\ s^{-1}}$,  $v_{th\ H}=14{\rm\ km\ s^{-1}}$.
Refer the text for the definitions of these parameters. 
This profile provides a much better fit than that shown in Fig.~6.
It is noted that the expansion velocity of the H~I shell is $v_{exp}=15
{\rm\ km\ s^{-1}}$, which is significantly smaller than that considered
in Fig.~6.
}
\label{final}
\end{figure}
%%% Table 1

\begin{deluxetable}{lccc}
\tablewidth{0pt}
\tabletypesize{\scriptsize}
\tablecaption{Single Gaussian Fit Parameters of Emission Lines}
\tablehead{
\colhead{Line} &
\colhead{$\lambda_0$\ (\AA)} &
\colhead{$f_0$} &
\colhead{$\Delta\lambda$\ (\AA)}} 
\startdata
H$\alpha$~6563                &  6563.23   &  34.8     &  0.54    \\
He~II~$\lambda$~6560          &  6560.58   &  0.072    &  0.48    \\
He~II~$\lambda$~6527          &  6527.49   &  0.00295  &  0.48    \\
${\rm [N~II]}\ \lambda$~6548  &  6548.51   &  0.897    &  0.47    \\
${\rm [N~II]}\ \lambda$~6583  &  6583.90   &  2.73     &  0.48    \\
\enddata
\end{deluxetable}
 
%%% Table 2

\begin{deluxetable}{lcc}
\tablewidth{0pt}
\tabletypesize{\scriptsize}
\tablecaption{He II Recombination Data by Storey \& Hummer (1995)}
\tablehead{
\colhead{ Line Ratio} &
\colhead{ $T_e=10^4{\rm\ K}$}        &
\colhead{ $T_e=2\times10^4{\rm\ K}$}}        
\startdata
{$n_e=10^4{\rm\ cm^{-3}}$} & & \\
$F_{1025}/F_{6560}$ &   3.600   &  4.519    \\
$F_{6527}/F_{6560}$ &   $3.952\times10^{-2}$   &  $4.085\times10^{-2}$    \\
\hline
{$n_e=10^6{\rm\ cm^{-3}}$} & & \\
$F_{1025}/F_{6560}$ &   3.804   &  4.676    \\
$F_{6527}/F_{6560}$ &   $4.098\times10^{-2}$   &  $4.152\times10^{-2}$    \\
\hline
{$n_e=10^8{\rm\ cm^{-3}}$} & & \\
$F_{1025}/F_{6560}$ &   4.439   &  5.181    \\
$F_{6527}/F_{6560}$ &   $4.942\times10^{-2}$   &  $4.614\times10^{-2}$    \\
\enddata
\end{deluxetable}


\begin{thebibliography}{}
\bibitem[Aller et al. (1996)]{all96} Aller, L. H., Hyung, S. \& Feibelman, W.,
A., 1996, \pasp, 108, 488
\bibitem[Altschuler et al. (1986)]{alt86} Altschuler, D. R., Schneider, S. E.,
Giovanardi, C., \& Silverglate, P. R., 1986, \apjl, 305, L85
\bibitem[Balick (1989)]{bal89} Balick, B., 1989, \aj, 97, 476
\bibitem[Birriel (2004)]{bir04} Birriel, J. J., 2004, \apj, 612, 1136
\bibitem[Corradi \& Scwarz 1995]{cor95} Corradi, R. \& Schwarz, H. E., 1995, 
\aap, 293, 871
\bibitem[Dinerstein et al. 1995]{din95} Dinerstein, H. L., Sneden, C., \&
Uglum, J., 1995, \apj, 447, 262
\bibitem[Espey (1995)]{espey95} Espey, B. R., Schulte-Ladbeck, R. E.,
Kriss, G. A., Hamann. F., Schmid, H. M., Johnson, J. J., 1995, \apj, 454, L61
\bibitem[Gathier et al.]{gat95} Gathier, R., Pottasch, S. R. \&
Goss, W. M., 1986, A\&A, 157, 191
\bibitem[Groves et al. (2002)]{gro02} Groves, B., Dopita, M. A.,
Williams, R. E., \& Hua, C. -T., 2002, PASA, 19, 425
\bibitem[Gruendl et al. 2004]{gru04} Gruendl, R. A., Chu, Y.-H., 
Guerrero, M. A., \apjl, 617, L127
\bibitem[Gussie and Taylor (1995)]{gus95} Gussie, G. T., \& Taylor, A. R.,
1995, \mnras, 273, 801
\bibitem[Harries et al. 1996]{har96} Harries, T. J., \& Howarth, I. D., 
1996, A\&AS, 119, 61

\bibitem[Ikeda et al. 2004]{ike04} Ikeda, Y., Akitaya, H., Matsuda, K.,
Homma, K., Seki, M., Kawabata, K. S., Hirata, R., Okazaki, A., 2004, \apj,
604, 357
\bibitem[Jung and Lee (2004a)]{jun04a} Jung, Y. -C., \& Lee, H. -W., 2004a,
\mnras, 350, 580
\bibitem[Jung and Lee (2004b)]{jun04b} Jung, Y. -C., \& Lee, H. -W., 2004b,
\mnras, 355, 221
\bibitem[Kim et al. (2007)]{kim07} Kim, K. -M., Han, I., Valyavin, G. G.,
Plachinda, S., Jang, J. G., Jang, B. -H., Seong, H. C., Lee, B. -C., Kang, D.
-I., Park, B. -G., Yoon, T. S., \& Vogt, S. S., 2007, \pasp, 119, 1052
\bibitem[Kwok et al. (2003)]{kwo03} Kwok, S., Su, K., Y. L., \& Sahai, R.,
2003, IAUS, 209, 481
\bibitem[Lee et al. (2006)]{lee06} Lee, H. -W., Jung, Y. -C., Song, I. -O.
\& Ahn, S. -H., 2006, \apj, 636, 1045
\bibitem[Lee and Kang (2007)]{lee07} Lee, H. -W., \& Kang, S., 2007, \apj, 
669, 1156
\bibitem[Lee et al. (2001)]{lee01} Lee, H. -W., Kang, Y. -W., Byun, Y., -I.,
2001, \apj, 551, L121
\bibitem[Lee and Park (1999)]{lee99} Lee, H. -W., \& Park, M.-G.,
1999, \apjl, 515, L89
\bibitem[Nussbaumer, Schmid \& Vogel, 1989]{nus89} Nussbaumer, H.,
Schmid, H. M., \& Vogel, M., 1989, \aap, 211, L27
\bibitem[Osterbrock 1987]{ost87} Osterbrock, D., 1987,
Astrophysics of Gaseous Nebulae and Active Galactic Nuclei, University Science
Books, Mill Valley
\bibitem[P\'equignot et al. (1997)]{peq97} P\'equignot, D., Baluteau, J. -P.,
Morisset, C., \& Boisson, C., 1997, \aap, 323, 217
\bibitem[Prinja et al. 2007]{pri07} Prinja, R. K., Hodges, S. E., Massa, D. L.,
Fullerton, A. W., Burnley, A. W., 2007, \mnras, 382, 299
\bibitem[Rybicki(1979)]{Ryb79} Rybicki, G. B., \& Lightman, A. P., 1979, 
Radiative Processes in Astrophysics, John Wiley \& Sons, Inc., New York
\bibitem[Schmid (1989)]{sch89} Schmid, H. M., 1989, \aap, 211, L31
\bibitem[Schmid (1998)]{sch98} Schmid, H. M., 1998,
Reviews in Modern Astronomy, 11, 297
\bibitem[Schneider et al. (1987)]{sch87} Schneider, S. E., Silver, P. R.,
Altschuler, D. R., \& Giovanardi, C., 1987, \apjl, 314, 572
\bibitem[Storey \& Hummer 1995]{sto95} Storey, P. J. \& Hummer, D. G.,
1995, \mnras, 272, 41
\bibitem[Taylor et al. (1990)]{tay90} Taylor, A. R., Gussie, G. T., \&
Pottasch, S. R., 1990, \apj, 351, 515
\bibitem[Tylenda et al (2003)]{tyl03} Tylenda, R., Si\'odmiak, N.,
G\'orny, S. K., Corradi, R. L. M., Schwarz, H. E., 2003, A\&A, 405, 627
\bibitem[Ueta et al. 2001]{uet01} Ueta, T., Fong, D. \& Meixner, M., 2001,
\apjl, 557, L117
\bibitem[Van Groningen (1993)]{van93} Van Groningen, E., 1993, \mnras,
264, 975
\bibitem[Yoo et al (2002)]{yoo02} Yoo, J. J., Bak, J.-Y., Lee, H. -W., 2002,
\mnras, 336, 467
\bibitem[Zhang 1995]{zha95} Zhang, C. Y., 1995, \apjs, 98, 659 
\end{thebibliography}
\end{document}